\documentclass[a4paper,11pt]{article}
\usepackage{pos}
 \usepackage[nice]{nicefrac}
\def\HP{\hphantom{\alpha}} 
\usepackage[utf8]{inputenc}
\usepackage{graphicx}
\usepackage{amsmath}
\usepackage{amsfonts}
\usepackage{physics}
\usepackage{xspace}
\usepackage{bm}
\usepackage{mathrsfs}
\usepackage{nicefrac}
\usepackage{multirow}
\usepackage[format=plain,justification=RaggedRight,labelsep=endash,font=small,labelfont=sc]{caption}
\usepackage{stackengine}

\usepackage[export]{adjustbox}

\def\GLW{{\rm GLW}}

\definecolor {darkgreen}{rgb}{0.2,0.7,0.2}

\def\be{\begin{equation}}
	\def\ee{\end{equation}}
\newcommand{\bel}[1]{\begin{eqnarray}\label{#1}}
	\newcommand{\eel}{\end{eqnarray}}
\def\barr{\begin{array}}
	\def\earr{\end{array}}
\def\beq{\begin{eqnarray}}
	\def\eeq{\end{eqnarray}}
\def\bfig{\begin{figure}}
	\def\efig{\end{figure}}
\def\lt{\left}
\def\rt{\right}
\def\CHI{\chi}
\newcommand{\nn}{\nonumber}

\newcommand{\p}{\partial}

\def\a{\alpha}
\def\b{\beta}
\def\g{\gamma}
\def\d{\delta} 
\def\r{\rho}
\def\s{\sigma}
\def\c{\chi}
 
\def\LR{\left(} 
\def\RR{\right)}
\def\LS{\left[} 
\def\RS{\right]}
\def\LC{\left{} 
\def\RC{\right}}
\def\LA{\left\langle}
\def\RA{\right\rangle}
\def\LD{\left.}
\def\RD{\right.}
\def\HP{\hphantom{\alpha}} 

\def\half{\frac{1}{2}}

\def\GLW{{\rm GLW}}
\def\LRFF{{\rm LRFF}}

\def\nU{n_{(0)}}
\def\eU{\varepsilon_{(0)}}
\def\PU{P_{(0)}}
\def\sU{s_{(0)}}

\def\nP{n_{}}
\def\eP{\varepsilon_{}}
\def\PP{P_{}}
\def\sP{s_{}}
\def\wP{w_{}}

\newcommand{\lab}[1]{\label{#1}}

\def\pmu{p^\mu}
\def\pnu{p^\nu}

\def\vv{{\boldsymbol v}}
\def\pv{{\boldsymbol p}}
\def\av{{\boldsymbol a}}
\def\bv{{\boldsymbol b}}
\def\kv{{\boldsymbol k}}
\def\omnL{\omega_{\mu\nu}}
\def\omnU{\omega^{\mu\nu}}
\def\omnLbar{{\bar \omega}_{\mu\nu}}
\def\omnUbar{{\bar \omega}^{\mu\nu}}
\def\omnLbardot{{\dot {\bar \omega}}_{\mu\nu}}
\def\omnUbardot{{\dot {\bar \omega}}^{\mu\nu}}

\def\oabL{\omega_{\alpha\beta}}
\def\oabU{\omega^{\alpha\beta}}
\def\omnLD{{\tilde \omega}_{\mu\nu}}
\def\omnUD{\tilde {\omega}^{\mu\nu}}
\def\omnLDbar{{\bar {\tilde \omega}}_{\mu\nu}}
\def\omnUDbar{{\bar {\tilde {\omega}}}^{\mu\nu}}

\def\epsLmnbg{\epsilon_{\mu\nu\beta\gamma}}
\def\epsUmnbg{\epsilon^{\mu\nu\beta\gamma}}
\def\epsLmnab{\epsilon_{\mu\nu\alpha\beta}}
\def\epsUmnab{\epsilon^{\mu\nu\alpha\beta}}

\def\epsUmnrs{\epsilon^{\mu\nu\rho \sigma}}
\def\epsUlnrs{\epsilon^{\lambda \nu\rho \sigma}}
\def\epsUlmrs{\epsilon^{\lambda \mu\rho \sigma}}

\def\epsLmnbg{\epsilon_{\mu\nu\beta\gamma}}
\def\epsUmnbg{\epsilon^{\mu\nu\beta\gamma}}
\def\epsLmnab{\epsilon_{\mu\nu\alpha\beta}}
\def\epsUmnab{\epsilon^{\mu\nu\alpha\beta}}

\def\epsLabgd{\epsilon_{\alpha\beta\gamma\delta}}
\def\epsUabgd{\epsilon^{\alpha\beta\gamma\delta}}

\def\epsUmnrs{\epsilon^{\mu\nu\rho \sigma}}
\def\epsUlnrs{\epsilon^{\lambda \nu\rho \sigma}}
\def\epsUlmrs{\epsilon^{\lambda \mu\rho \sigma}}

\def\epsLijk{\epsilon_{ijk}}
\def\half{\frac{1}{2}}

\def\GLW{{\rm GLW}}

\def\n0{n_{(0)}}
\def\e0{\varepsilon_{(0)}}
\def\P0{P_{(0)}}
\title{Dissipative effects on the
propagation of spin modes}

\author*[a,b]{Rajeev Singh}
\affiliation[a]{Center for Nuclear Theory, Department of Physics and Astronomy, Stony Brook University, Stony Brook, New York, 11794-3800, USA}
\affiliation[b]{Department of Modern Physics, University of Science and Technology of China, Hefei, Anhui 230026, China}
\emailAdd{rajeevofficial24@gmail.com}
\author[c,d]{Victor E. Ambrus}
	\emailAdd{victor.ambrus@e-uvt.ro}
    \affiliation[c]{
Institut f\"ur Theoretische Physik, Johann Wolfgang Goethe-Universit\"at, Max-von-Laue-Strasse 1, D-60438 Frankfurt am Main, Germany}%
\affiliation[d]{Department of Physics, West University of Timisoara,
Bd.~Vasile P\^arvan 4, Timisoara 300223, Romania}
\author[e]{Radoslaw Ryblewski}
\emailAdd{radoslaw.ryblewski@ifj.edu.pl}
\affiliation[e]{Institute of Nuclear Physics Polish Academy of Sciences, PL 31-342 Krak\'ow, Poland}

\abstract{In relativistic hydrodynamics with spin, following de Groot--van Leeuwen--van Weert's energy-momentum and spin tensor definitions, we analyze the propagation of spin degrees of freedom. We deduce an analytical formula for spin wave velocity, finding that it approaches half the speed of light in the ultra-relativistic limit. Only transverse degrees of freedom propagate, similar to electromagnetic waves. Additionally, we explore dissipative effects and determine the damping coefficients for Maxwell-J\"uttner statistics.}

\FullConference{%
  25th International Spin Symposium (SPIN 2023)\\
  24-29 September, 2023}

\begin{document}

\def\be{\begin{equation}}
	\def\ee{\end{equation}}
	\def\barr{\begin{array}}
	\def\earr{\end{array}}
\def\beq{\begin{eqnarray}}
	\def\eeq{\end{eqnarray}}
\def\bfig{\begin{figure}}
	\def\efig{\end{figure}}
\newcommand{\bea}{\begin{eqnarray}}
	\newcommand{\eea}{\end{eqnarray}}

\def\LB{\left(}
\def\RB{\right)}
\def\LSB{\left[}
\def\RSB{\right]}
\def\LAB{\langle}
\def\RAB{\rangle}

\newcommand{\VP}{\vphantom{\frac{}{}}\!}
\def\lt{\left}
\def\rt{\right}
\def\CHI{\chi}
\newcommand{\rfmtwo}[2]{Eqs.~(\ref{#1})-(\ref{#2})}
\newcommand{\rfcs}[1]{Refs.~\cite{#1}}

\def\a{\alpha}
\def\b{\beta}
\def\g{\gamma}
\def\d{\delta} 
\def\r{\rho}
\def\s{\sigma}
\def\c{\chi} 
 \def\lam{\lambda} 
\def\LR{\left(} 
\def\RR{\right)}
\def\LS{\left[} 
\def\RS{\right]}
\def\LC{\left{} 
\def\RC{\right}}
\def\LA{\left\langle}
\def\RA{\right\rangle}
\def\LD{\left.}
\def\RD{\right.}
\def\HP{\hphantom{\alpha}} 



\def\half{\frac{1}{2}}

\def\GLW{{\rm GLW}}
\def\LRFF{{\rm LRFF}}


\def\nU{n_{(0)}}
\def\nUi{n_{(0),i}}
\def\eU{\varepsilon_{(0)}}
\def\eUi{\varepsilon_{(0),i}}
\def\PU{P_{(0)}}
\def\PUi{P_{(0),i}}
\def\sU{s_{(0)}}
\def\sU{s_{(0),i}}

\def\nP{n_{}}
\def\eP{\varepsilon_{}}
\def\PP{P_{}}
\def\sP{s_{}}
\def\wP{w_{}}

\def\nn{\nonumber}



\def\cA{{\cal A}}
\def\cB{{\cal B}}
\def\cC{{\cal C}}
\def\cD{{\cal D}}
\def\cN{{\cal N}}
\def\cE{{\cal E}}
\def\cP{{\cal P}}
\def\cS{{\cal S}}
\def\cT{{\cal T}}
\def\cQ{{\cal Q}}
\def\cNN{{\cal N}_{(0)}}
\def\cEN{{\cal E}_{(0)}}
\def\cPN{{\cal P}_{(0)}}
\def\cSN{{\cal S}_{(0)}}


\def\pmu{p^\mu}
\def\pnu{p^\nu}

\def\vv{{\boldsymbol v}}
\def\pv{{\boldsymbol p}}
\def\av{{\boldsymbol a}}
\def\bv{{\boldsymbol b}}
\def\kv{{\boldsymbol k}}
\def\omnL{\omega_{\mu\nu}}
\def\omnU{\omega^{\mu\nu}}
\def\omnLbar{{\bar \omega}_{\mu\nu}}
\def\omnUbar{{\bar \omega}^{\mu\nu}}
\def\omnLbardot{{\dot {\bar \omega}}_{\mu\nu}}
\def\omnUbardot{{\dot {\bar \omega}}^{\mu\nu}}

\def\oabL{\omega_{\alpha\beta}}
\def\oabU{\omega^{\alpha\beta}}
\def\omnLD{{\tilde \omega}_{\mu\nu}}
\def\omnUD{\tilde {\omega}^{\mu\nu}}
\def\omnLDbar{{\bar {\tilde \omega}}_{\mu\nu}}
\def\omnUDbar{{\bar {\tilde {\omega}}}^{\mu\nu}}
\def\CHI{\chi}
\def\bchem{\mu_{\rm B}}
\def\bfug{\xi_{\rm B}}
\def\tfug{\xi}

\def\be{\begin{equation}}
\def\ee{\end{equation}}
\def\ba{\begin{eqnarray}}
\def\ea{\end{eqnarray}}   

\def\a{\alpha}
\def\b{\beta}
\def\g{\gamma}
\def\d{\delta} 
\def\r{\rho}
\def\s{\sigma}
\def\c{\chi}
 
\def\LR{\left(} 
\def\RR{\right)}
\def\LS{\left[} 
\def\RS{\right]}
\def\LC{\left{} 
\def\RC{\right}}
\def\LA{\left\langle}
\def\RA{\right\rangle}
\def\LD{\left.}
\def\RD{\right.}
\def\half{\frac{1}{2}}

\def\GLW{{\rm GLW}}
\def\LRF{{\rm LRF}}


\def\nU{n_{(0)}}
\def\eU{\varepsilon_{(0)}}
\def\PU{P_{(0)}}
\def\sU{s_{(0)}}

\def\nP{n_{}}
\def\eP{\varepsilon_{}}
\def\PP{P_{}}
\def\sP{s_{}}
\def\wP{w_{}}


\def\pmu{p^\mu}
\def\pnu{p^\nu}

\def\vv{{\boldsymbol v}}
\def\pv{{\boldsymbol p}}
\def\av{{\boldsymbol a}}
\def\bv{{\boldsymbol b}}
\def\cv{{\boldsymbol c}}
\def\Cv{{\boldsymbol C}}
\def\kv{{\boldsymbol k}}
\def\piv{{\boldsymbol \pi}}

\def\thetap{\theta_\perp}
\def\omnL{\omega_{\mu\nu}}
\def\omnU{\omega^{\mu\nu}}
\def\omnLbar{{\bar \omega}_{\mu\nu}}
\def\omnUbar{{\bar \omega}^{\mu\nu}}
\def\omnLbardot{{\dot {\bar \omega}}_{\mu\nu}}
\def\omnUbardot{{\dot {\bar \omega}}^{\mu\nu}}

\def\oabL{\omega_{\alpha\beta}}
\def\oabU{\omega^{\alpha\beta}}
\def\omnLD{{\tilde \omega}_{\mu\nu}}
\def\omnUD{\tilde {\omega}^{\mu\nu}}
\def\omnLDbar{{\bar {\tilde \omega}}_{\mu\nu}}
\def\omnUDbar{{\bar {\tilde {\omega}}}^{\mu\nu}}

\def\epsLmnbg{\epsilon_{\mu\nu\beta\gamma}}
\def\epsUmnbg{\epsilon^{\mu\nu\beta\gamma}}
\def\epsLmnab{\epsilon_{\mu\nu\alpha\beta}}
\def\epsUmnab{\epsilon^{\mu\nu\alpha\beta}}

\def\epsUmnrs{\epsilon^{\mu\nu\rho \sigma}}
\def\epsUlnrs{\epsilon^{\lambda \nu\rho \sigma}}
\def\epsUlmrs{\epsilon^{\lambda \mu\rho \sigma}}

\def\epsLmnbg{\epsilon_{\mu\nu\beta\gamma}}
\def\epsUmnbg{\epsilon^{\mu\nu\beta\gamma}}
\def\epsLmnab{\epsilon_{\mu\nu\alpha\beta}}
\def\epsUmnab{\epsilon^{\mu\nu\alpha\beta}}

\def\epsLabgd{\epsilon_{\alpha\beta\gamma\delta}}
\def\epsUabgd{\epsilon^{\alpha\beta\gamma\delta}}

\def\epsUmnrs{\epsilon^{\mu\nu\rho \sigma}}
\def\epsUlnrs{\epsilon^{\lambda \nu\rho \sigma}}
\def\epsUlmrs{\epsilon^{\lambda \mu\rho \sigma}}

\def\epsLijk{\epsilon_{ijk}}


\def\epsLmnbg{\epsilon_{\mu\nu\beta\gamma}}
\def\epsUmnbg{\epsilon^{\mu\nu\beta\gamma}}
\def\epsLmnab{\epsilon_{\mu\nu\alpha\beta}}
\def\epsUmnab{\epsilon^{\mu\nu\alpha\beta}}

\def\epsUmnrs{\epsilon^{\mu\nu\rho \sigma}}
\def\epsUlnrs{\epsilon^{\lambda \nu\rho \sigma}}
\def\epsUlmrs{\epsilon^{\lambda \mu\rho \sigma}}

\def\epsLmnbg{\epsilon_{\mu\nu\beta\gamma}}
\def\epsUmnbg{\epsilon^{\mu\nu\beta\gamma}}
\def\epsLmnab{\epsilon_{\mu\nu\alpha\beta}}
\def\epsUmnab{\epsilon^{\mu\nu\alpha\beta}}

\def\epsLabgd{\epsilon_{\alpha\beta\gamma\delta}}
\def\epsUabgd{\epsilon^{\alpha\beta\gamma\delta}}

\def\epsUmnrs{\epsilon^{\mu\nu\rho \sigma}}
\def\epsUlnrs{\epsilon^{\lambda \nu\rho \sigma}}
\def\epsUlmrs{\epsilon^{\lambda \mu\rho \sigma}}

\def\epsLijk{\epsilon_{ijk}}
\def\half{\frac{1}{2}}
\def\GLW{{\rm GLW}}

\def\n0{n_{(0)}}
\def\e0{\varepsilon_{(0)}}
\def\P0{P_{(0)}}
\newcommand{\redflag}[1]{{\color{red} #1}}
\newcommand{\blueflag}[1]{{\color{blue} #1}}
\newcommand{\checked}[1]{{\color{darkblue} \bf [#1]}}
\newcommand{\Psis}{{\sf \Psi}}
\newcommand{\psis}{{\sf \psi}}
\newcommand{\Psibar}{{\overline \Psi}}
\def\eMf{electromagnetic (EM) }
\def\EMf{Electromagnetic (EM) }
\def\EM{EM }
\def\lRFf{local rest frame (LRF)}
\def\LRFf{Local rest frame (LRF) }
\def\LRF{LRF }
\def\QGPf{Quark gluon plasma (QGP) }
\def\qGPf{Quark gluon plasma (QGP) }
\def\QGP{QGP }
\def\mHDf{magnetohydrodynamic (MHD) }
\def\MHDf{Magnetohydrodynamic (MHD) }
\def\MHD{MHD }
\def\iMHD{iMHD }
\def\HD{Hydrodynamics }
\def\hD{hydrodynamics }
\def\RHD{Relativistic hydrodynamics }
\def\rHD{relativistic hydrodynamics }
\def\rMHDf{relativistic magnetohydrodynamic (RMHD) }
\def\RMHDf{Relativistic magnetohydrodynamic (RMHD) }
\def\RMHD{RMHD }
\def\eOMf{equations of motion (EOM)~}
\def\EOMf{Equations of motion (EOM)~}
\def\EOM{EOM}
\def\fl{\ensuremath{\text{Fluid}}}
\def\lrf{\ensuremath{\text{LRF}}}
\def\BVf{Boltzmann-Vlasov (BV) }
\def\BV{BV\,}
		
\def\rhoLEQ{{\widehat{\rho}}_{\rm \small LEQ}}
\def\rhoGEQ{{\widehat{\rho}}_{\rm \small GEQ}}
		
\def\fplushat{{\hat f}^+}
\def\fminushat{{\hat f}^-}
		
\def\fplusrs{f^+_{rs}}
\def\fplussr{f^+_{sr}}
\def\fplusrsxp{f^+_{rs}(x,p)}
\def\fplussrxp{f^+_{sr}(x,p)}
		
\def\fminusrs{f^-_{rs}}
\def\fminussr{f^-_{sr}}
\def\fminusrsxp{f^-_{rs}(x,p)}
\def\fminussrxp{f^-_{sr}(x,p)}
		
\def\fpmrs{f^\pm_{rs}}
\def\fpmrsxp{f^\pm_{rs}(x,p)}

\def\feqplus{f^+_{eq}}
\def\feqplus{f^+_{eq}}
\def\feqplusxp{f^+_{eq}(x,p)}
\def\feqplusxp{f^+_{eq}(x,p)}
		
\def\feqminus{f^-_{eq}}
\def\feqminus{f^-_{eq}}
\def\feqminusxp{f^-_{eq}(x,p)}
\def\feqminusxp{f^-_{eq}(x,p)}
	
\def\feqpm{f^\pm_{{\rm eq}}}
\def\feqpmxp{f^\pm_{{\rm eq}}(x,p)}
\def\feqpmi{f^\pm_{{\rm eq},i}}
\def\feqpmxpi{f^\pm_{{\rm eq},i}(x,p)}
\def\fpm{f^\pm}
\def\fpmxp{f^\pm(x,p)}
\def\fpmi{f^\pm_i}
\def\fpmxpi{f^\pm_i(x,p)}
\newcommand{\rs}[1]{\textcolor{red}{#1}}
\newcommand{\rrin}[1]{\textcolor{blue}{#1}}
\newcommand{\rrout}[1]{\textcolor{blue}{\sout{#1}}}
\newcommand{\lie}[2]{\pounds_{#1}\,#2}
\newcommand{\rd}{\mathrm{d}}
\def\re{\mathrm{e}}
\def\echarge{\ensuremath{\rho_e}}
\def\cond{\ensuremath{{\sigma_e}}}
\newcommand{\msnote}[1]{\todo[author=Masoud]{#1}}
\newcommand{\msnotei}[1]{\todo[author=Masoud,inline]{#1}}
\newcommand{\explainindetail}[1]{\todo[color=red!40]{#1}}
\newcommand{\insertref}[1]{\todo[color=green!40]{#1}}
\newcommand{\fm}{\rm{\,fm}}
\newcommand{\fmc}{\rm{\,fm/c}}


\def\uv{{\boldsymbol U}}


\def\kbarzero{ {\bar k}^0}
\def\kv{{\boldsymbol k}}
\def\kbarv{{\bar {\boldsymbol k}}}

\def\obarzero{ {\bar \omega}^0}
\def\ov{{\boldsymbol \omega}}
\def\obar{{\bar \omega}}
\def\obarv{{\bar {\boldsymbol \omega}}}

\def\ev{{\boldsymbol e}}
\def\bv{{\boldsymbol b}}
\newcommand{\tT}{\theta_T}
\newcommand{\UD}[1]{\oU{#1}}
\newcommand{\XD}[1]{\oX{#1}}
\newcommand{\YD}[1]{\oY{#1}}
\newcommand{\ZD}[1]{\oZ{#1}}
\newcommand\oU[1]{\ensurestackMath{\stackon[1pt]{#1}{\mkern2mu\bullet}}}
\newcommand\oX[1]{\ensurestackMath{\stackon[1pt]{#1}{\mkern2mu\star}}}
\newcommand\oY[1]{\ensurestackMath{\stackon[1pt]{#1}{\mkern2mu\smwhitestar}}}
\newcommand\oZ[1]{\ensurestackMath{\stackon[1pt]{#1}{\mkern2mu\circ}}}
\def\Aone{{ \cal A}_1 }
\def\Atwo{{ \cal A}_2 }
\def\Athree{{ \cal A}_3 }
\def\Afour{{ \cal A}_4 }
\def\vv{{\boldsymbol v}}
\def\pv{{\boldsymbol p}}

\newcommand{\inv}[1]{\frac{1}{#1}}
\newcommand{\iinv}[1]{1/#1}
\maketitle
\section{Introduction}
\label{sec:introduction}
%
Measurements of $\Lambda(\Bar{\Lambda})$ hyperon spin polarization~\cite{STAR:2017ckg,STAR:2019erd,ALICE:2019aid,ALICE:2019onw} have ignited significant interest in spin dynamics, particularly concerning spin-orbit coupling~\cite{Liang:2004ph,Liang:2004xn,Gao:2007bc,Chen:2008wh,Wang:2023csj,Xie:2023gbo}. The polarization is theoretically rooted in spin-orbit coupling as per the Dirac equation, with Vilenkin's 1980s work~\cite{Vilenkin:1980zv} indicating a chiral flow along vorticity in a Dirac particle gas~\cite{Kharzeev:2015znc}. This chiral flow, linked to the axial anomaly and affected by vorticity or electromagnetic (EM) fields, is termed `anomalous transport'~\cite{Kharzeev:2015znc} and led to hydrodynamics with triangle anomalies~\cite{Son:2009tf}. However, for massive particles like hyperons, axial current conservation is explicitly broken, making the axial chemical potential model less suitable~\cite{Chernodub:2020yaf}.
Models based on spin thermodynamic equilibrium~\cite{Becattini:2017gcx,Becattini:2021suc,Becattini:2021iol,Fu:2021pok} have aligned well with polarization data, but differential polarization measurements remain unclear~\cite{STAR:2019erd,ALICE:2021pzu}. This spurred the integration of spin into hydrodynamics~\cite{Florkowski:2017ruc,Florkowski:2017dyn}, using energy-momentum and spin tensor definitions by de Groot, van Leeuwen, and van Weert (GLW)~\cite{Florkowski:2018ahw}, with recent developments in this formalism~\cite{Florkowski:2018ahw,Florkowski:2019qdp,Bhadury:2020puc,Singh:2020rht,Singh:2021man,Florkowski:2021wvk,Ambrus:2022yzz,Singh:2022ltu,Singh:2022uyy}.
This paper analyzes linear perturbations in perfect-fluid spin hydrodynamics~\cite{Florkowski:2018fap,Florkowski:2018ahw}, finding that spin degrees of freedom in the conservation equation are decoupled from the background fluid, leading to separate analyses for fluid and spin wave spectra~\cite{Ambrus:2017keg}. The spin tensor linearization implies validity only for unpolarized backgrounds, yielding a general analytic spin wave velocity expression both Maxwell-Jüttner (MJ), revealing $c_{\rm spin} = c/2$ in the ultra-relativistic limit. Spin degrees of freedom are divided into electric ($C_{\boldsymbol{\kappa}}$) and magnetic ($C_{\boldsymbol{\omega}}$) parts, analogous to EM waves. 
An analysis of the dissipative part of the spin tensor~\cite{Bhadury:2020puc,Bhadury:2020cop} for the ideal MJ gas shows that dissipative effects lead to the damping of both the transverse and the longitudinal components of the spin tensor~\cite{Ambrus:2022yzz}.
This paper is structured as follows:\footnote{We adopt the Minkowski metric $g_{\mu\nu} = \hbox{diag}(+1,-1,-1,-1)$ and natural (Planck) units, with $c = \hbar = k_B~=1$ (unless stated otherwise). The dot product of four-vectors $a^{\alpha}$ and $b^{\alpha}$ is $a \cdot b = a^{\alpha}b_{\alpha} = a^0 b^0 - \av \cdot \bv$, and for the Levi-Civita tensor $\epsilon^{\alpha\beta\gamma\delta}$, we use $\epsilon^{txyz} = +1$. Antisymmetrization is denoted by square brackets, $M_{[\mu \nu]} = \frac{1}{2}(M_{\mu\nu} - M_{\nu\mu})$.}
a review of spin hydrodynamics is presented in Sec.~\ref{sec:spinhydro}, followed by an analysis of
spin polarization perturbations and wave solutions in Sec.~\ref{sec:wave_analysis}.
Dissipation effects on
the propagation of the
spin wave
are discussed in Sec.~\ref{sec:diss}. Section~\ref{sec:conclusion} summarizes our conclusions.
\section{Relativistic spin hydrodynamics}
\label{sec:spinhydro}
This section summarizes the GLW-based hydrodynamic framework for spin-$\frac{1}{2}$ particles with mass $m$, where spin effects are small, not influencing the conservation laws for charge, energy, and momentum, but stemming from angular momentum conservation. The conservation laws for baryon current and energy-momentum tensor are outlined as~\cite{Florkowski:2017ruc,Florkowski:2018ahw,Florkowski:2018fap}.
\ba
\p_\alpha N^\alpha(x)  = 0\,, \quad \p_\b T^{\a\b}(x) = 0\,,
\lab{Ncon}
\ea
where $N^\alpha$ and $T^{\alpha\beta}$ are the baryon current and the energy-momentum tensor, respectively. For a perfect (non-dissipative) fluid, we have~\cite{Florkowski:2017ruc}
\ba
N^\alpha = {\cal N} U^\alpha\,, \quad 
T^{\a\b} = {\cal E} U^\a U^\b - 
{\cal P} ~\Delta^{\a\b}\,.
\lab{Nmu}
\ea
In this context, ${\cal N}$ represents the baryon charge density, ${\cal E}$ the energy density, and ${\cal P}$ the pressure. The fluid's four-velocity is indicated by $U^{\mu}$, and $\Delta^{\a\b} = g^{\a\b} - U^\a U^\b$ serves as the projector onto the surface orthogonal to $U^{\mu}$.
The symmetry of the energy-momentum tensor (\ref{Nmu}) necessitates separate conservation of spin, as per total angular momentum conservation,
$\p_\a S^{\a , \beta \gamma }(x)= 0$~\cite{Florkowski:2018ahw}.

Quantum effects like non-local collisions can lead to deviations from above conservation equation~\cite{Hidaka:2018ekt,Weickgenannt:2020aaf,Yang:2020hri,Wang:2020pej,Weickgenannt:2021cuo}, potentially causing the spin polarization tensor $\omega^{\mu\nu}$ \eqref{spinpol1} to align with the local thermal vorticity. However, as the precise relaxation equation is still unknown, such effects are not included in our current analysis. At leading order, the spin tensor is decomposed as~\cite{Florkowski:2018ahw,Florkowski:2018fap,Florkowski:2021wvk}.
\begin{subequations}\label{eq:S}
\beq
S^{\alpha , \beta \gamma }
&=&  S^{\alpha , \beta \gamma }_{\rm ph} + S^{\a, \b\g}_{\Delta},
\label{eq:SGLW}
\eeq
where $S^{\alpha , \beta \gamma }_{\rm ph}$ (phenomenological)
and $S^{\a, \b\g}_{\Delta}$ (auxiliary) contributions are~\cite{Florkowski:2017ruc,Florkowski:2021wvk}
\beq
S^{\alpha , \beta \gamma }_{\rm ph}
&=&  (\mathcal{A}_1 + \mathcal{A}_3) U^\alpha \omega^{\beta\gamma},\label{Spheno}\\
S^{\a, \b\g}_{\Delta} 
&=&  (2\mathcal{A}_1-\mathcal{A}_3) 
\, U^\a U^\d U^{[\b} \omega^{\g]}_{\HP\d} 
+ \mathcal{A}_3 \left( 
\Delta^{\a\d} U^{[\b} \omega^{\g]}_{\HP\d}+ U^\a \Delta^{\d[\b} \omega^{\g]}_{\HP\d}
+ U^\d \Delta^{\a[\b} \omega^{\g]}_{\HP\d}\right),
\lab{eq:SDeltaGLW} 
\eeq
\end{subequations}
with the thermodynamic coefficients~\cite{Ambrus:2022yzz}
\beq
\mathcal{A}_1 = \frac{\mathbf{\mathfrak{s}}^2}{9}\left[ \left(\frac{\partial \mathcal{N}}{\partial \xi}\right)_{\beta} - \frac{2}{m^2} \left( \frac{\partial \mathcal{E}}{\partial \beta}
\right)_{\xi}
\right],\quad 
\mathcal{A}_3 = \frac{2\mathbf{\mathfrak{s}}^2}{9}\left[ \left(\frac{\partial \mathcal{N}}{\partial \xi}\right)_{\beta} + \frac{1}{m^2} \left( \frac{\partial \mathcal{E}}{\partial \beta}
\right)_{\xi}
\right].
\label{eq:A1&A3}
\eeq
We utilized general formulas for $\mathcal{A}_1$ and $\mathcal{A}_3$ that are statistics-independent in our kinetic model~\cite{Ambrus:2022yzz}. For expressions specific to MJ statistics in an ideal gas, see Ref.~\cite{Florkowski:2018ahw}. Here, $\xi = \mu / T$ denotes the chemical potential to temperature ratio, $\beta$ is the 
inverse
temperature, and $\mathbf{\mathfrak{s}}^2 = s(s+1)$, which equals $3/4$ for spin-$\frac{1}{2}$ particles, represents the
square of the
spin angular momentum magnitude~\cite{Florkowski:2018fap}. Additionally, we define $z = m /T$ as the mass to temperature ratio.
The antisymmetric spin polarization tensor $\omega_{\mu\nu}$ is given as~\cite{Florkowski:2017ruc}
\beq
\omega_{\mu\nu} &=& \kappa_\mu U_\nu - \kappa_\nu U_\mu + \epsilon_{\mu\nu\a\b} U^\a \omega^{\b}. \lab{spinpol1}
\eeq
In this structure, $\kappa_\mu$ and $\omega_\mu$ comprise six independent components. By design, these four-vectors are orthogonal to $U^{\mu}$, satisfying $\kappa_\mu~U^\mu = \omega_\mu~U^\mu = 0$,
\beq
\kappa_\mu= \omega_{\mu\a} U^\a, \quad \omega_\mu = \half \epsilon_{\mu\a\b\gamma} \omega^{\a\b} U^\gamma, \lab{eq:kappaomega}
\eeq
which, in the rest frame of the fluid,
reduce to
\beq
\kappa^\mu= \left(0, C_{\boldsymbol{\kappa}}\right) = (0, C_{\kappa X}, C_{\kappa Y}, C_{\kappa Z}), \quad \omega^\mu = \left(0, C_{\boldsymbol{\omega}}\right) = (0, C_{\omega X}, C_{\omega Y}, C_{\omega Z})\,, \lab{eq:kappaomega1}
\eeq
with $C_{\boldsymbol{\kappa}}$ and $C_{\boldsymbol{\omega}}$ being the spin polarization components~\cite{Florkowski:2018fap,Florkowski:2018ahw}.
\section{Spin mode dispersion relation} 
\label{sec:wave_analysis}
Now, examining the propagation of small disturbances in a fluid with spin degrees of freedom, we note that the background fluid's conservation equations (\ref{Ncon}) are unaffected by polarization~\cite{Florkowski:2018fap}, yielding the familiar sound wave spectrum~\cite{Ambrus:2017keg} that travels at the sound speed
\begin{eqnarray}
 c_s^2 = \left(\frac{\partial \mathcal{P}}{\partial \mathcal{E}}\right)_{\mathcal{N}} + 
 \frac{\mathcal{N}}{\mathcal{E} + \mathcal{P}} 
 \left(\frac{\partial \mathcal{P}}{\partial \mathcal{N}}\right)_{\mathcal{E}}.
\end{eqnarray}
Turning to excitations within the spin tensor \eqref{eq:SGLW}, we assume a stationary background fluid ($U^\mu = g^{t\mu}$) and treat $\omega^{\mu\nu}$ as a small factor, indicating an unpolarized background fluid. Under these conditions, Eqs.~\eqref{Spheno} and \eqref{eq:SDeltaGLW} simplify to
\begin{eqnarray}
 S^{\alpha,\mu\nu}_{\rm ph} &=& (\mathcal{A}_1 + \mathcal{A}_3) g^{t\alpha} \omega^{\mu\nu},\nonumber\\
 S^{\alpha,\mu\nu}_{\Delta} &=& 2(\mathcal{A}_1 - 2\mathcal{A}_3)
 g^{t\alpha} g^{t[\mu} \omega^{\nu]t}  + \mathcal{A}_3 (g^{t[\mu} \omega^{\nu] \alpha} + 
 g^{\alpha [\mu} \omega^{\nu]t} - 
 g^{t\alpha} \omega^{\mu\nu}).
 \label{eq:S_lin}
\end{eqnarray}
Assuming homogeneity in the $x$ and $y$ directions, taking the divergence of Eq.~\eqref{eq:S_lin} results in
\begin{eqnarray}
 \partial_\alpha S^{\alpha,\mu\nu}_{\rm ph} &=& (\mathcal{A}_1  + \mathcal{A}_3) \partial_t \omega^{\mu\nu},\nonumber\\
 \partial_\alpha S^{\alpha,\mu\nu}_{\Delta} &=& 
 (2\mathcal{A}_1 - 3\mathcal{A}_3) g^{t[\mu} \partial_t \omega^{\nu] t}  + \mathcal{A}_3 (\partial^{[\mu} \omega^{\nu] t} - 
 \partial_t \omega^{\mu\nu} + g^{t[\mu} \partial_z \omega^{\nu] z}),
\end{eqnarray}
which for $\mu = 0, \nu = i$ and $\mu = i, \nu = j$, respectively, give
\begin{eqnarray}
 \partial_\alpha S^{\alpha,ti} = \mathcal{A}_3\left(\partial_t \omega^{ti} + \frac{1}{2} \partial_z \omega^{i z}\right), \quad
 \partial_\alpha S^{\alpha,ij} = \mathcal{A}_1 
 \partial_t \omega^{ij} + \mathcal{A}_3 \partial^{[i} \omega^{j] t}.
 \label{eq:dSaux}
\end{eqnarray}
\begin{figure}[t]
\includegraphics[width=0.495\columnwidth]{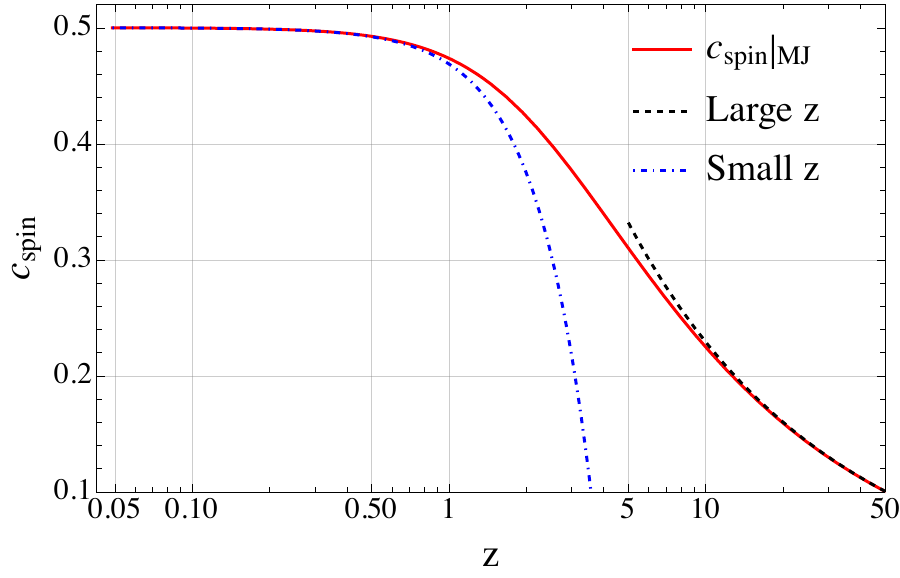}
\includegraphics[width=0.495\columnwidth]{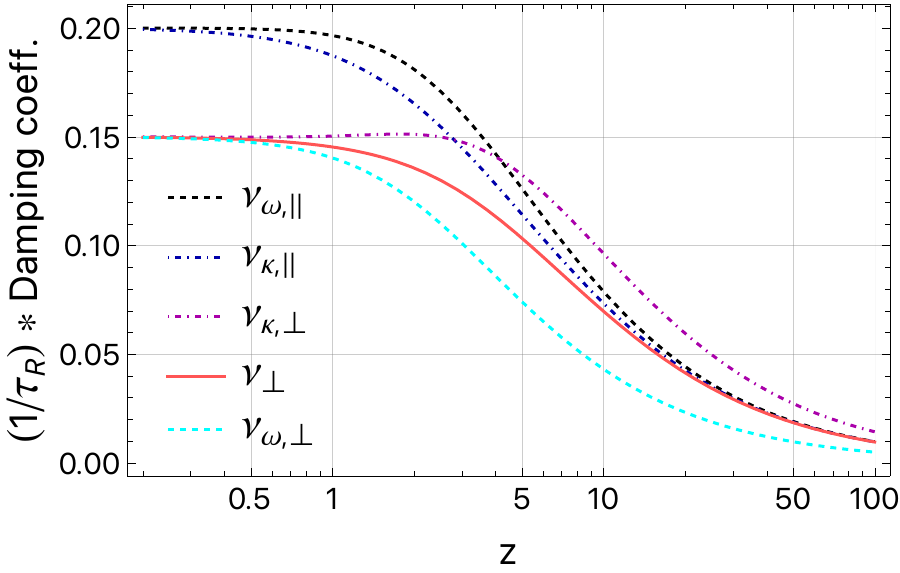}
\caption{(Left panel) $c_{\rm spin}$ as a function of $z$ 
for MJ statistics together with small and large $z$ limits~\cite{Singh:2022uyy}.
(Right panel) The damping coefficients' dependence on $z$ for longitudinal ($\nu_{\omega,||}$, $\nu_{\kappa,||}$, beginning at $0.2$) and transverse ($\nu_{\kappa,\perp}$, $\nu_\perp$, and $\nu_{\omega,\perp}$, starting at $0.15$) modes, calculated with $\mathbf{\mathfrak{s}}^2 = 3/4$ as per Eqs.~\eqref{eq:diss_nu}, \eqref{eq:nulong}, and \eqref{eq:nuperp}~\cite{Ambrus:2022yzz}.}
\label{fig:cspin}
\end{figure}
Considering Eq.~\eqref{eq:kappaomega1}, the components of $\omega^{\mu\nu}$ are expressible in terms of $C_{\kappa i}$ and $C_{\omega k}$ as
\begin{eqnarray}
 \omega^{ti} = -C_{\kappa i}, \qquad 
 \omega^{ij} = -\epsilon^{tijk} C_{\omega k}.
\end{eqnarray}
Requiring $\partial_\alpha S^{\alpha,\mu\nu} = 0$, we get
\begin{eqnarray}
 \partial_t C_{\kappa i} - \frac{1}{2}\epsilon^{tijz} \partial_z C_{\omega j} 
 = 0, \quad
 \partial_t C_{\omega i} - 
 \frac{\mathcal{A}_3}{2\mathcal{A}_1} 
 \epsilon^{tijz} \partial_z C_{\kappa j} = 0.
 ~~~~~~~\label{eq:eqC}
\end{eqnarray}
The presence of the Levi-Civita symbol results in $\partial_t C_{\kappa Z} = \partial_t C_{\omega Z} = 0$, meaning that 
the longitudinal components do not propagate. Consequently, polarization propagates solely as transverse waves, akin to EM waves. This is derived by 
considering
$i = x, y$ in Eq.~\eqref{eq:eqC}, yielding
\begin{eqnarray}
 \left(\frac{\partial^2}{\partial t^2} - c_{\rm spin}^2 \frac{\partial^2}{\partial z^2}\right) \mathcal{C} = 0,
\end{eqnarray}
with $\mathcal{C} \in \{C_{\kappa X}, C_{\kappa Y}, C_{\omega X}, C_{\omega Y}\}$ and spin wave speed fulfils
\begin{eqnarray}
 c_{\rm spin}^2 = -\frac{1}{4} \frac{\mathcal{A}_3}
 {\mathcal{A}_1} = 
 \frac{1}{4} \frac{(\partial \mathcal{E} / \partial T)_\xi - 
 z^2 (\partial \mathcal{N} / \partial \xi)_T} 
 {(\partial \mathcal{E} / \partial T)_\xi + 
 \frac{z^2}{2} (\partial \mathcal{N} / \partial \xi)_T}.
 \label{eq:cspin}
\end{eqnarray}
In $z \rightarrow 0$ (ultra-relativistic) limit, 
$c_{\rm spin}$ reduces to 
$1/2$ regardless of statistics. 
The formula for $c_{\rm spin}^2$ can be written for the ideal MJ gas as
\begin{eqnarray}
 c_{\rm spin}^2 = \frac{1}{4} \frac{K_3(z)}{K_3(z) + \frac{z}{2} K_2(z)},\label{eq:cspin_Kideal}
\end{eqnarray}
that is independent of chemical potential $\xi$, which for $z\ll 1$ and for $z \gg 1$ (non-relativistic limit) reduce to
\begin{align}
 c_{\rm spin}\Big|_{z\ll 1} = \frac{1}{2} 
 \left[1 - \frac{z^2}{16} + O(z^4)\right], \quad c_{\rm spin}\Big|_{z\gg 1} \simeq \frac{1}{\sqrt{2z}}\,,
 \label{eq:cspin_smallz}
\end{align}
%
%
respectively. Figure~\ref{fig:cspin} shows that $c_{\rm spin}$ monotonically decreases with increasing $z$ as $0 < c_{\rm spin} \le \frac{1}{2}$.
The lower limit of this range corresponds to a cold, massive particle gas (non-relativistic limit), and the upper limit applies to high temperatures or massless particles.
\section{Dissipative effects on spin mode propagation}
\label{sec:diss}
This section examines how dissipation affects the propagation of spin modes, drawing on the analysis of dissipative effects from Ref.~\cite{Bhadury:2020cop}.
Under the relaxation time approximation, the dissipative adjustments to $T^{\mu\nu}$ and $N^\mu$ are found to be independent of the spin tensor, with the dissipative correction to the spin component expressed as
\begin{equation}
 \delta S^{\lambda,\mu\nu}= \tau_R (B^{\lambda,\mu\nu}_\Pi \theta + 
 B^{\kappa\lambda,\mu\nu}_n \nabla_\kappa \xi + 
 B^{\kappa\delta\lambda,\mu\nu}_\pi \sigma_{\kappa\delta} 
 + B^{\eta\beta\gamma\lambda,\mu\nu}_\Sigma 
 \nabla_\eta \omega_{\beta\gamma}),
 \label{eq:Bhadury}
\end{equation}
with $\tau_R$ being the relaxation time and $\nabla_\mu = \Delta_\mu{}^\nu \partial_\nu = \partial_\mu - U_\mu U^\nu \partial_\nu.$
In this section, we continue examining small perturbations in an unpolarized background in thermal equilibrium. Our focus is on understanding how dissipation affects these perturbations in spin hydrodynamics and ensuring the theory's stability. We treat perturbation amplitudes, including $\delta S^{\lambda,\mu\nu} \sim \omega^{\mu\nu}$, as infinitesimal but allow for large gradient magnitudes proportional to the wave number $k$. This approach helps identify potential instabilities at small wavelengths ($k \rightarrow \infty$), though it's important to note that our analysis may not fully capture the physics when $\tau_R \gg 1$ and/or $k \gg 1$.
Referring to Eq.~\eqref{eq:Bhadury}, the coefficients $B^{\lambda,\mu\nu}_\Pi$, $B^{\kappa\lambda,\mu\nu}_n$, and $B^{\kappa\delta\lambda,\mu\nu}_\pi$ are linked to the spin polarization tensor $\omega^{\mu\nu}$, treated as first-order relative to perturbation amplitude in an unpolarized background. These coefficients are multiplied by first-order gradient terms like $\theta = \partial_\mu u^\mu$, $\nabla_\kappa \xi$, and $\sigma_{\kappa\delta} = \frac{1}{2}(\nabla_\kappa u_\delta + \nabla_\delta u_\kappa) - \frac{1}{3} \theta \Delta_{\kappa\delta}$. Given their second-order nature with respect to perturbation amplitude, the first three terms in Eq.~\eqref{eq:Bhadury} can be disregarded, allowing us to concentrate on the last term~\cite{Bhadury:2020cop}
\begin{eqnarray}
 && B^{\eta\beta\gamma\lambda,\mu\nu}_\Sigma = 
 B^{(1)}_\Sigma \Delta^{\lambda\eta} g^{\beta[\mu} g^{\nu] \gamma} + 
 B^{(2)}_\Sigma \Delta^{\lambda\eta} u^\gamma u^{[\mu} \Delta^{\nu]\beta}\\
 &&+ B^{(3)}_\Sigma(\Delta^{\lambda\eta} \Delta^{\gamma[\mu} g^{\nu]\beta} + 
 \Delta^{\lambda\gamma} \Delta^{\eta[\mu} g^{\nu] \beta} + 
 \Delta^{\gamma\eta} \Delta^{\lambda[\mu} g^{\nu] \beta})
 + B^{(4)}_\Sigma \Delta^{\gamma\eta} \Delta^{\lambda[\mu} \Delta^{\nu] \beta} + 
 B^{(5)}_\Sigma u^\gamma \Delta^{\lambda\beta} u^{[\mu} \Delta^{\nu]\eta},\nonumber
\end{eqnarray}
where $B^{(i)}_\Sigma$ are~\cite{Ambrus:2022yzz}
\begin{align}
 B^{(1)}_\Sigma =& -\frac{4 \mathbf{\mathfrak{s}}^2 \cosh \xi}{3} I^{(1)}_{21}, 
 \quad
  B^{(2)}_\Sigma = -\frac{8 \mathbf{\mathfrak{s}}^2 \cosh \xi}{3m^2}  \left(I^{(1)}_{41} + \frac{I^{(1)}_{41} I^{(0)}_{31}}
  {m^2 I^{(0)}_{10} - 2 I^{(0)}_{31}}\right), \quad B^{(3)}_\Sigma = -\frac{8 \mathbf{\mathfrak{s}}^2 \cosh \xi}{3m^2} I^{(1)}_{42},
  \nonumber\\
  B^{(4)}_\Sigma =& -\frac{8 \mathbf{\mathfrak{s}}^2 \cosh \xi}{3m^2}  
  \frac{I^{(1)}_{41} I^{(0)}_{31}}{m^2 I^{(0)}_{10} - (I^{(0)}_{30} 
  + I^{(0)}_{31})},\quad
  B^{(5)}_\Sigma = \frac{8 \mathbf{\mathfrak{s}}^2 \cosh \xi}{3m^2} 
  \frac{I^{(1)}_{41} I^{(0)}_{31}}{m^2 I^{(0)}_{10} - 2I^{(0)}_{31}},
\end{align}
and $I^{(r)}_{nq}$ are thermodynamic integrals.
%
As $A^\eta \nabla_\eta \omega_{\beta\gamma}$ simplifies to $A^z \partial_z \omega_{\beta\gamma}$, the index $\eta$ can be effectively replaced with $z$, hence, we proceed with the splitting
\begin{eqnarray}
 \partial_\lambda \delta S^{\lambda,\mu\nu} = \tau_R \sum_i B_\Sigma^{(i)} 
 \mathfrak{T}^{(i) \mu\nu},
\end{eqnarray}
and obtain 
\begin{align}
 \mathfrak{T}^{(1)\mu\nu} =& -\partial_z^2 \omega^{\mu\nu}, \quad
 \mathfrak{T}^{(2)\mu\nu} = -g^{t[\mu} \partial_z^2 \omega^{\nu]t}, \quad
 \mathfrak{T}^{(3)\mu\nu} = \partial_z^2 \omega^{\mu\nu} + 
 g^{t[\mu} \partial^2_z \omega^{\nu]t} + 
 2g^{z[\mu} \partial^2_z \omega^{\nu] z},\nonumber\\\
 \mathfrak{T}^{(4)\mu\nu} =& g^{z[\mu} \partial^2_z \omega^{\nu]z} - g^{z[\mu} g^{\nu]t} \partial^2_z \omega^{tz}, \quad
 \mathfrak{T}^{(5)\mu\nu} = g^{z[\mu} g^{\nu]t} \partial^2_z \omega^{tz}.
\end{align}
Summing the above terms together, we have
\begin{eqnarray}
 \frac{1}{\tau_R} \partial_\lambda \delta S^{\lambda,\mu\nu} &=& 
 -\LR B^{(1)}_\Sigma - B^{(3)}_\Sigma\RR \partial_z^2 \omega^{\mu\nu}
 - \LR B^{(2)}_\Sigma - B^{(3)}_\Sigma\RR g^{t[\mu} \partial^2_z \omega^{\nu] t}
 + 2B^{(3)}_\Sigma g^{z[\mu} \partial^2_z \omega^{\nu] z}\nonumber \\
 &-&\LR B^{(4)}_\Sigma - B^{(5)}_\Sigma\RR 
 g^{z[\mu} g^{\nu] t} \partial^2_z \omega^{tz} + 
 B^{(4)}_\Sigma g^{z[\mu} \partial^2_z \omega^{\nu] z}.
\end{eqnarray}
We know that $\omega^{ti} = -C_{\kappa i}$ and $\omega^{ij} = -\epsilon^{0ijk} C_{\omega k}$, thus, we get
\begin{align}
 \partial_\lambda \delta S^{\lambda, t x} =& \nu_{\kappa,\perp} \mathcal{A}_3 
 \partial^2_z C_{\kappa X}, &
 \partial_\lambda \delta S^{\lambda, yz} =& 
 \nu_{\omega,\perp} \mathcal{A}_1 \partial^2_z C_{\omega X}, \nonumber\\
 \partial_\lambda \delta S^{\lambda, t y} =& \nu_{\kappa,\perp} \mathcal{A}_3 
 \partial^2_z C_{\kappa Y}, & 
 \partial_\lambda \delta S^{\lambda, zx} =& 
 \nu_{\omega,\perp} \mathcal{A}_1 \partial^2_z C_{\omega Y},
 \nonumber\\
 \partial_\lambda \delta S^{\lambda, tz} =& \nu_{\kappa, ||} \mathcal{A}_3
 \partial^2_z C_{\kappa Z}, & 
 \partial_\lambda \delta S^{\lambda, xy} =& 
 \nu_{\omega,||} \mathcal{A}_1 \partial^2_z C_{\omega Z},
 \label{eq:diss_eqs_aux}
\end{align}
where we identified ($\nu_{\kappa,||}$, $\nu_{\omega,||}$) and ($\nu_{\kappa,\perp}$, $\nu_{\omega,\perp}$) as longitudinal  and transverse 
kinematic viscosities, respectively,
\begin{align}
 \nu_{\kappa, ||} =& \frac{\tau_R}{\mathcal{A}_3} B^{(3)}_\Sigma,\quad
 \nu_{\omega,||} = \frac{\tau_R}{\mathcal{A}_1} \left(B^{(1)}_\Sigma - B^{(3)}_\Sigma\right),\nonumber\\
 \nu_{\kappa,\perp} =& \frac{\tau_R}{\mathcal{A}_3} \left(B^{(1)}_\Sigma - \frac{1}{2} B^{(2)}_\Sigma - 
 \frac{1}{2} B^{(3)}_\Sigma\right),\quad
 \nu_{\omega,\perp} = \frac{\tau_R}{\mathcal{A}_1} 
 \left(B^{(1)}_\Sigma - 2 B^{(3)}_\Sigma - 
 \frac{1}{2} B^{(4)}_\Sigma \right).
 \label{eq:diss_nu}
\end{align}
Putting in Eq.~\eqref{eq:dSaux} reveals that both $C_{\kappa Z}$ and $C_{\omega Z}$ demonstrate exponential decay
\begin{eqnarray}
 \partial_t C_{\kappa Z} - \nu_{\kappa,||} \partial^2_z C_{\kappa Z} = 0,\quad
 \partial_t C_{\omega Z} - \nu_{\omega,||} \partial^2_z C_{\omega Z} = 0.
\end{eqnarray}
Assuming $C_{\kappa/\omega; Z}\!\! \sim \!\! e^{-i \omega t + i k z} \widetilde{C}_{\kappa/\omega; Z}$, 
where $\widetilde{C}_{\kappa/\omega; Z}$ is a constant, we obtain $\omega = -i k^2 \nu_{\kappa/\omega, ||}$ where
\begin{eqnarray}
 \nu_{\kappa,||} &=& \frac{4\mathbf{\mathfrak{s}}^2 \tau_R}{45 G(z)} 
 [-5z + G(z)(3 + z^2) - z^2 {\rm Gi}(z)] \simeq \frac{4 \mathbf{\mathfrak{s}}^2 \tau_R}{15}
 \left[1 - \frac{z^2}{12} + O(z^4)\right],\nonumber\\
 \nu_{\omega,||} &=& \frac{4\mathbf{\mathfrak{s}}^2 \tau_R}{15(2G(z) + z)} 
 [5z + G(z)(2 - z^2) + z^2 {\rm Gi}(z)] \simeq \frac{4 \mathbf{\mathfrak{s}}^2 \tau_R}{15}
 \left[1 - \frac{z^4}{16} + O(z^5)\right],
 \label{eq:nulong}
\end{eqnarray}
with $G(z) = K_3(z) / K_2(z)$ and ${\rm Gi}(z) = {\rm Ki}_1(z) / K_2(z)$.
Further employing a
Fourier decomposition 
$\mathcal{C} = \widetilde{\mathcal{C}} e^{-i \omega t + i k z}$ of the transverse modes give
\beq
 \begin{pmatrix}
  \omega + i k^2 \nu_{\kappa, \perp} & -k/2\\
  \frac{k \mathcal{A}_3}{2 \mathcal{A}_1} & \omega + i k^2 \nu_{\omega,\perp} 
 \end{pmatrix}
 \begin{pmatrix}
 \widetilde{C}_{\kappa X} \\
 \widetilde{C}_{\omega Y} 
 \end{pmatrix} = 0, \quad
 \begin{pmatrix}
  \omega + i k^2 \nu_{\kappa,\perp} & k/2 \\
  - \frac{k \mathcal{A}_3}{2 \mathcal{A}_1} & 
  \omega + i k^2 \nu_{\omega,\perp}
 \end{pmatrix}
 \begin{pmatrix}
 \widetilde{C}_{\kappa Y} \\
 \widetilde{C}_{\omega X}
 \end{pmatrix} = 0,
\eeq
where the dispersion relation is
\begin{align}
 \omega_\pm = -i k^2 \nu_\perp \pm k c_{\rm spin}\,, \qquad 
 \nu_\perp = \frac{\nu_{\kappa,\perp} + \nu_{\omega,\perp}}{2}\,,
 \quad
 c_{\rm spin}^2 = -\frac{\mathcal{A}_3}{4\mathcal{A}_1} - \frac{k^2}{4} \left(\nu_{\kappa,\perp} - \nu_{\omega,\perp}\right)^2.
 \label{eq:diss_disp}
\end{align}
The damping coefficient, $\nu_\perp$, is the average of the individual damping coefficients for $\kappa$ and $\omega$
\begin{align}
 \nu_\perp = \frac{2 \mathbf{\mathfrak{s}}^2 \tau_R [3G(z) + 2z]}{45G(z)[2G(z) + z]} 
 [- 5z + G(z)(3 + z^2) - z^2 {\rm Gi}(z)]
 \simeq \frac{\mathbf{\mathfrak{s}}^2 \tau_R}{5} 
 \left[1 - \frac{z^2}{24} + O(z^4)\right].
 \label{eq:nuperp}
\end{align}
The spin wave speed undergoes a negative dissipative correction, estimated as $c_{\rm spin}^2 = c^2_{\rm spin; 0}(1 - \delta c^2_{\rm spin})$, with $c_{\rm spin;0}^2 = -\mathcal{A}_3 / 4 \mathcal{A}_1 > 0$, and
\begin{equation}
 \delta c^2_{\rm spin} = \frac{k^2 (\nu_{\kappa,\perp} - \nu_{\omega,\perp})^2}{-\mathcal{A}_3 / \mathcal{A}_1}
 \simeq \frac{k^2 \mathbf{\mathfrak{s}}^4 \tau_R^2}{8100} z^4[1 + O(z^6)],
\end{equation}
which is suppressed for small values of $z$.
However, for finite $z$, a sufficiently large wave number can make $c_{\rm spin}^2$ negative, occurring when $k$ surpasses a certain threshold value
\begin{eqnarray}
k_{\rm th} = \frac{2 c_{\rm spin;0}}{|\nu_{\kappa,\perp} - \nu_{\omega,\perp}|}.
\end{eqnarray}
When $k$ exceeds $k_{\rm th}$, $c_{\rm spin}$ turns imaginary, halting wave propagation, similar to effects seen in first-order hydrodynamics of spinless systems. An example is sound modes in ultra-relativistic fluids, where $\tau_R k_{\rm th} = \frac{5\eta}{4\mathcal{P}} k_{\rm th} = 15/2$~\cite{Ambrus:2017keg}. In scenarios where $k \gg k_{\rm th}$, Eq.~\eqref{eq:diss_disp} indicates that stability is maintained provided,
$\nu_\perp - \frac{1}{2} |\nu_{\kappa,\perp} - \nu_{\omega,\perp} |  = 
 {\rm min}(\nu_{\kappa,\perp}, \nu_{\omega,\perp}) > 0$,
which remains true for the formalism studied here.
This can be checked analytically for small values of $z$, when
\begin{eqnarray}
 \nu_{\kappa,\perp} \simeq \frac{\mathfrak{s}^2 \tau_R}{5}\left[1 - \frac{z^2}{72} + O(z^4)
 \right],\quad
 \nu_{\omega,\perp} \simeq \frac{\mathfrak{s}^2 \tau_R}{5}\left[1 - \frac{5z^2}{72} + O(z^4)
 \right],
\end{eqnarray}
and $\tau_R k_{\rm th} \simeq 18 / (5 z^2 \mathfrak{s}^2)$.
The right panel of Fig.~\ref{fig:cspin} shows that both $\nu_{\kappa,\perp}$ and $\nu_{\omega,\perp}$ stay positive for large $z$, indicating stability against linear perturbations.

We now assess 
the effect of dissipation
on spin wave propagation in heavy-ion collisions, focusing on the $z \ll 1$ 
limit. In this case,
the shear viscosity $\eta$ relates to the relaxation time as $\eta = \frac{4}{5} \tau_R \mathcal{P}$~\cite{Ambrus:2017keg}. 
Imposing
a constant $\eta / \mathcal{S}$ ratio, where $\mathcal{S} = (\mathcal{E} + \mathcal{P} - \mu \mathcal{N}) / T \approx 4\mathcal{P} / T$ represents the entropy density (assuming $|\xi| \ll 1$), we find
\begin{eqnarray}
 \tau_R \simeq \frac{5}{4\pi^2 T} \times (4\pi \eta /\mathcal{S}).
\end{eqnarray}
Putting $\mathfrak{s}^2 = 3/4$, the damping time 
$t_{\rm damp; \perp} = 1 / k^2 \nu_\perp$ can be computed as
\begin{align}
 t_{\rm damp; \perp} \simeq \frac{4\lambda^2 T/3}{4\pi \eta / \mathcal{S}} 
 = \left(\frac{\lambda}{1\ {\rm fm}}\right)^2 
 \left(\frac{T}{600\ {\rm MeV}}\right) \times 
 \frac{4\ {\rm fm} / c}{4\pi \eta / \mathcal{S}},
\end{align}
with $\lambda = 2\pi / k$ as the wavelength, indicating
that the lifespan of spin waves is comparable to that of the QGP fireball.
\section{Summary}
\label{sec:conclusion}
In this study, we explored the wave spectrum in spin hydrodynamics using the GLW pseudogauge, focusing on the antisymmetric tensor $\omega^{\mu\nu}$ with six independent degrees of freedom, split into three electric and three magnetic components. Our findings highlight the transverse nature of spin waves in ideal fluids, where longitudinal components don't propagate, but transverse ones do, similar to EM waves.
The spin wave speed, $c_{\rm spin}$, varies with medium parameters (temperature $T$, chemical potential $\mu$) and particle properties (mass $m$, statistics). In the ultra-relativistic limit ($z = m/T \ll 1$), $c_{\rm spin} \simeq 1/2$, regardless of statistics. For an ideal MJ gas, $c_{\rm spin}$ is unaffected by $\xi = \mu /T$. In the large $z$ limit, we found that $c_{\rm spin} \sim 1/\sqrt{2z}$ for classical (Maxwell-J\"uttner) statistics.
Dissipation effects on spin waves show that all transverse components are damped similarly ($\nu_\perp$), while longitudinal components decay at different rates ($\nu_{\kappa,||}$ and $\nu_{\omega,||}$). Viscous corrections affect $c_{\rm spin}$ significantly at high wave numbers, turning imaginary beyond 
a threshold wavenumber $k_{\rm th}$, thus
preventing wave propagation.
This approach, focusing on $\omega^{\alpha\beta}$, does not encompass anomalous transport phenomena. Adding vortical terms to $N^\alpha$ and $T^{\alpha\beta}$ alters the fluid sector's wave spectrum, introducing excitations like the chiral magnetic wave \cite{Kharzeev:2010gd}, chiral vortical wave \cite{Jiang:2015cva}, or helical vortical wave \cite{Ambrus:2019khr}. Future research could intriguingly explore the interaction between anomalous transport and spin polarization tensor dynamics.

\smallskip
{\it Acknowledgements.}
R.S. acknowledges the support of Polish NAWA Bekker program No.: BPN/BEK/2021/1/00342.
V.E.A. acknowledges support through a grant of the Ministry of Research, Innovation and Digitization, CNCS - UEFISCDI, project number PN-III-P1-1.1-TE-2021-1707, within PNCDI III.
This research was also supported in part by the Polish National Science Centre Grant No. 2018/30/E/ST2/00432.
\bibliographystyle{utphys}
\bibliography{fluctuationRef.bib}{}
\end{document}